# Synthesis and superconductivity of new $BiS_2$ based superconductor $PrO_{0.5}F_{0.5}BiS_2$

Rajveer Jha, Anuj Kumar, Shiva Kumar Singh, and V.P. S. Awana

Quantum Phenomena and Applications Division, National Physical Laboratory (*CSIR*), Dr. K.S. Krishnan Marg, New Delhi-110012, India

We report synthesis and superconductivity at 3.7K in $PrO_{0.5}F_{0.5}BiS_2$. The newly discovered material belongs to the layered sulfide based $REO_{0.5}F_{0.5}BiS_2$ compounds having ZrCuSiAs type structure. The bulk polycrystalline compound is synthesized by vacuum encapsulation technique at $780^0C$ in single step. Detailed structural analysis has shown that the as synthesized $PrO_{0.5}F_{0.5}BiS_2$ is crystallized in tetragonal *P4/nmm* space group with lattice parameters $a = 4.015(5)$ Å, $c = 13.362(4)$ Å. Bulk superconductivity is observed in $PrO_{0.5}F_{0.5}BiS_2$ below 4K from magnetic and transport measurements. Electrical transport measurements showed superconducting transition temperature ($T_c$) onset at 3.7K and $T_c$ ($\rho=0$) at 3.1K. Hump at $T_c$ related to superconducting transition is not observed in heat capacity measurement and rather a Schottky-type anomaly is observed at below ~6K. The compound is slightly semiconducting in normal state. Isothermal magnetization (*MH*) exhibited typical type II behavior with lower critical field ($H_{c1}$) of around 8Oe.

*Corresponding Author
Dr. V. P. S. Awana, Senior Scientist
E-mail: awana@mail.npindia.org
Ph. +91-11-45609357, Fax-+91-11-45609310
Homepage www.fteewebs.com/vpsawana/

Recent reports on superconductivity of layered $Bi_4O_4S_3$ [1-5] and $(La/Nd/Ce)O_{0.5}F_{0.5}BiS_2$ [6-10] had already attracted a lot of attention. These compounds are similar in terms of their layered structure and carrier doping mechanism to popular high $T_c$ cuprates and Fe-pnictides. Perhaps the role of $Cu-O_2$ planes of cuprates and FeAs of pnictides is being played by $BiS_2$ layers in these compounds. For normal pressure synthesized samples the superconducting transition temperature is reported around 4K for $Bi_4O_4S_3$ [1-5], 2.7K for $LaO_{0.5}F_{0.5}BiS_2$ [6,7], 5K for $NdO_{0.5}F_{0.5}BiS_2$ [8,9] and at 4K for $CeO_{0.5}F_{0.5}BiS_2$ [10]. Electronic structure calculations showed that these compounds doping mechanism could be the same as for Fe pnictides and their parent un-doped phase is non superconducting [11-13]. The $BiS_2$ based layered superconductivity is yet in preliminary stage, still at least one systematic on-site substitution study is reported on suppression of superconductivity by Bi site Ag doping in $Bi_4O_4S_3$ [14]. Also reported is a new compound i.e., $SrFBiS_2$, which seems to be the ground state of these $BiS_2$ based layered superconductors [15]. In fact appropriate doping of mobile carriers in $SrFBiS_2$ could lead to higher $T_c$. In this short note we report the first synthesis and superconductivity of nearly phase pure new member i.e., $PrO_{0.5}F_{0.5}BiS_2$, belonging to the same family of reported $(La/Nd/Ce)O_{0.5}F_{0.5}BiS_2$ [6-10] compounds with popular ZrCuSiAs type structure. The as synthesized $PrO_{0.5}F_{0.5}BiS_2$ compound is nearly single phase in nature with small contamination of $Bi_2S_3$ and bulk superconducting below 4K as established by both electrical transport and magnetization measurements.

Polycrystalline $PrO_{0.5}F_{0.5}BiS_2$ is synthesized by standard solid state reaction route via vacuum encapsulation. High purity Pr, Bi, S, $PrF_3$, and $Pr_6O_{11}$ are weighed in stoichiometric ratio and ground thoroughly in a glove box under high purity argon atmosphere. The mixed powders are subsequently palletized and vacuum-sealed ($10^{-4}$ Torr) in a quartz tube. Sealed quartz ampoule is placed in furnace and heat treated at $780^0C$ for 12h with the typical heating rate of $2^oC/min.$, and subsequently cooled down slowly over a span of six hours to room temperature. Thus processed compound was repelletized and sealed in quartz tube and heat treated in similar way. X-ray diffraction (*XRD*) was performed at room temperature using *Rigaku Diffractometer* with *Cu* $K_\alpha$ ($\lambda$ = 1.54Å). Rietveld analysis was carried out using the *FullProf* program. Detailed electrical transport, *AC/DC* magnetization and Heat capacity measurements were performed on Physical Property Measurements System (*PPMS*-14T), *Quantum Design*.

The room temperature Rietveld fitted *XRD* pattern of as synthesized $PrO_{0.5}F_{0.5}BiS_2$ sample is shown in Fig. 1. Rietveld refinement of *XRD* pattern is carried out using ZrCuSiAs structure and Wyckoff positions. The compound is crystallized in tetragonal *P4/nmm* space group structure. Small amount of $Bi_2S_3$ impurity is also observed along with main phase and is indicated in Fig. 1. Refined structural parameters (atomic coordinates and site occupancy), are given in the Table I. The refined lattice parameters are $a$ = 4.015(5) Å, $c$ = 13.362(4) Å. The representative unit cell of the compound in *P4/nmm* space group crystallization is shown in inset of Fig. 1. The layered structure includes rare earth oxide (*RE*O) and $BiS_2$ layers. The doping in *RE*O layer with proper amount of F⁻ at $O^{2-}$ site results in doping of carriers along with decrease in *c*-parameter [6,7]. Various atoms with their respective positions are indicated in the inset part of Fig. 1. Bismuth (Bi), Praseodymium (Pr), and Sulfur (S1 and S2) atoms occupy the *2c* (0.25, 0.25, *z*) site. On the other hand O/F atoms are at *2a* (0.75, 0.25, 0) site.

Both *DC* and *AC* magnetic susceptibility results are shown in Fig. 2. *DC* magnetization is carried out in both *ZFC* and *FC* protocols in an applied field of 5*Oe*. The compound exhibits sharp diamagnetic transition at around 3.7K in. The lower inset of figure shows the isothermal magnetization (*MH*) of the compound at 2.1K. $PrO_{0.5}F_{0.5}BiS_2$ clearly exhibits the type II behavior with its lower critical field ($H_{c1}$) at around 8Oe.

Fig. 3 presents the resistivity versus temperature (*ρ*-T) plots for $PrO_{0.5}F_{0.5}BiS_2$ sample in applied magnetic field of up to 4kOe in temperature range of 2-4K. Inset of the same shows the resistivity in expanded temperature range of 2-300K. The compound is semi-metallic from 300K down to 4K. The superconducting onset is observed in resistivity at 3.7K and the zero resistivity $T_c$ ($ρ = 0$) at 3.1K. The semi-metallic normal state conduction is observed which is similar to the reported results for $(La/Nd/Ce)O_{0.5}F_{0.5}BiS_2$ [6-10] compounds. With applied magnetic field $T_c(ρ = 0)$ is decreased to lower temperatures. Namely the $T_c(ρ = 0)$ of 3.1K in zero applied field decreases to 2.1K under 2.0kOe field.

Fig. 4 shows the d*ρ*/d*T* plots for the studied $PrO_{0.5}F_{0.5}BiS_2$ sample. Single characteristic peak is seen in d*ρ*/d*T* plots at various fields. Further, the peak temperature decreases with increase in applied field, notably in a single step. This is suggestive of single superconducting transition and better grains coupling in this system. The broadening of d*ρ*/d*T* peak under applied field suggests that superconducting onset is relatively affected less than the $T_c(ρ = 0)$ state. The peak temperature of d*ρ*/d*T* plots roughly corresponds with the superconducting state of the compound. The superconductivity sets in at onset of *ρ*-T plots and is complete at $ρ = 0$. The peak of d*ρ*/d*T* plots corresponds to the middle of superconductivity onset and exact $T_c(ρ = 0)$ state. In any case, with the application of magnetic field both the onset and offset $T_c$ shift towards lower temperature. The upper critical field $H_{c2}(T)$ and irreversibility field $H_{irr}(T)$ for $PrO_{0.5}F_{0.5}BiS_2$ can be calculated through the intersection point between normal state resistivity extrapolation and superconducting transition line respectively. The slope $dH_{c2}/dT|_{Tc}$ as estimated from Fig. 4 data is ~ 30*kOe/K* that implies *WHH* (Werthamer-Helfand-Hohenberg) $H_{c2}(0)$ (= - 0.69 $T_c$ $dH_{c2}/dT|_{Tc}$) value of 64*k*Oe, which is higher than the one (19*k*Oe) reported for $LaO_{0.5}F_{0.5}BiS_2$ of [7] and (23*k*Oe) for $NdO_{0.5}F_{0.5}BiS_2$ [9].

Heat capacity [$C_p$(T)] of $PrO_{0.5}F_{0.5}BiS_2$ is shown in Fig. 5. Hump at $T_c$ related to superconducting transition is not observed in heat capacity measurements and rather a Schottky-type anomaly is observed at below ~6K. This may be due to some incipient magnetic ordering, which may be dominating over superconducting transition of $C_p$. Similar type of Schottky anomaly is also observed for Nd/$CeO_{0.5}F_{0.5}BiS_2$ compound [9,10]. More rigorous experimental and theoretical studies are warranted to invoke these compounds. Also a hump like weak jump in $C_p$ measurements is observed for $LaO_{0.5}F_{0.5}BiS_2$ [6] and for $Bi_4O_4S_3$ [16] confirming the bulk nature of superconductivity in these $BiS_2$ based compounds. The upper inset (a) of Fig. 5 represents $C_p/T$ vs T plot and the lower inset (b) shows the $C_p/T$ vs $T^2$ plot. We have carried out the linear fitting of $C_p/T$ vs $T^2$ curve. It is estimated that the value of Sommerfeld constant (γ) and the coefficient of the lattice contribution (β), as 286.36 *mJ/mole-$K^2$* and 0.764 *mJ/mole-$K^4$* respectively.

In conclusion we have synthesized a near single phase $PrO_{0.5}F_{0.5}BiS_2$ compound, which is bulk superconducting at 3.7K. The compound is a type II superconductor with lower [$H_{c1}$ (2.1K)] of around 8Oe. This is first report on appearance of superconductivity in $PrO_{0.5}F_{0.5}BiS_2$ in the series of $BiS_2$ based layered *RE*O/F$BiS_2$ compounds.

1. Y. Mizuguchi, H Fujihisa, Y. Gotoh, , K. Suzuki, H. Usui, K. Kuroki, S. Demura, Y. Takano, H. Izawa, O. Miura, *arXiv:1207.3145 (2012)*: Phys. Rev. B. Accepted (2012).
2. S. K. Singh, A. Kumar, B. Gahtori, Shruti, G. Sharma, S. Patnaik, V. P. S. Awana, *arXiv:1207.5428 (2012)*: J. Am. Chem. Soc 134, 16504 (2012)
3. S. G. Tan, L. J. Li, Y. Liu, P. Tong, B. C. Zhao, W. J. Lu, and Y. P. Sun: *arXiv:1207.5395 (2012)*
4. S. Li, H. Yang, J. Tao, X. Ding, and H.-H. Wen: *arXiv:1207.4955 (2012)*
5. H. Kotegawa, Y. Tomita, H. Tou, H. Izawa, Y. Mizuguchi, O. Miura, S. Demura, K. Deguchi, and Y. Takano: *arXiv:1207.6935 (2012)*: J. Phys. Soc. Jpn. 81, 103702 (2012).
6. Y. Mizuguchi, S. Demura, K. Deguchi, Y. Takano, H. Fujihisa,Y. Gotoh, H. Izawa, O. Miura: *arXiv:1207.3558. (2012)*: J. Phys. Soc. Jpn. 81, 114525 (2012).
7. V. P. S. Awana, A. Kumar, R. Jha, S. K. Singh, J. Kumar, A. Pal, Shruti, J. Saha, and S. Patnaik: *arXiv:1207.6845 (2012)*: Solid State Commun. Accepted (2012).

8. S. Demura, Y. Mizuguchi, K. Deguchi, H.Okazaki, H. Hara, T. Watanabe, S. J. Denholme, M. Fujioka, T. Ozaki, H. Fujihisa, Y. Gotoh, O. Miura, T. Yamaguchi, H. Takeya, Y. Takano, arXiv:1207.5248 (2012)
9. Rajveer Jha, Anuj Kumar, Shiva Kumar Singh and V. P. S. Awana arXiv:1208.3077 (2012).
10. Jie Xing, Sheng Li, Xiaxing Ding, Huang Yang, and Hai-Hu Wen, 1208.5000 (2012).
11. H. Usui, K. Suzuki, K. Kuroki, arXiv:1207.3888 (2012)
12. Xiangang Wan, Hang-Chen Ding, Sergey Y. Savrasov, Chun-Gang Duan, arXiv:1208.1807 (2012)
13. Tao Zhou, Z. D. Wang arXiv:1208.1101 (2012)
14. S. G. Tan, P. Tong, Y. Liu, W. J. Ju, L. J. Li, B. C. Zhao and Y. P. Sun, arXiv:1208.5307 (2012)
15. Hechang Lei, Kefeng Wang, Milinda Abeykoon, Emil S. Bozin, and C. Petrovic, arXiv:1208.3189 (2012)
16. H. Takatsu, Y. Mizuguchi, H. Izawa, O. Miura, H. Kadowaki, J. Phys. Soc. Jap. DOI:10.1143/JPSJ.81.125002

**Table 1** Atomic coordinates, Wyckoff positions, and site occupancy for studied $PrO_{0.5}F_{0.5}BiS_2$.

| Atom | x | y | z | site | Occupancy |
|---|---|---|---|---|---|
| Pr | 0.250 | 0.250 | 0.098(4) | *2c* | 1 |
| Bi | 0.250 | 0.250 | 0.623(2) | *2c* | 1 |
| S1 | 0.250 | 0.250 | 0.372(1) | *2c* | 1 |
| S2 | 0.250 | 0.250 | 0.822(4) | *2c* | 1 |
| O  | 0.750 | 0.250 | 0.000 | *2a* | 0.5 |
| F  | 0.750 | 0.250 | 0.000 | *2a* | 0.5 |

**Figure Captions**

**Figure 1:** Observed (*open circles*) and calculated (*solid lines*) XRD patterns of $PrO_{0.5}F_{0.5}BiS_2$ compound at room temperature, inset shows the schematic unit cell of the compound.

**Figure 2:** AC magnetic susceptibility and *DC* magnetization (both *ZFC* and *FC)* plots for $PrO_{0.5}F_{0.5}BiS_2$. Inset shows isothermal *MH* curve of the sample at 2.1K, marking its lower critical field ($H_{c1}$).

**Figure 3:** Extended resistivity versus temperature ($\rho$-T) plots for $PrO_{0.5}F_{0.5}BiS_2$ sample, inset shows the same in 2-300K temperature range.

**Figure 4:** Variation of the derivative of resistivity with temperature (d$\rho$/dT) for $PrO_{0.5}F_{0.5}BiS_2$ sample.

**Figure 5:** Specific heat $C_p$ versus temperature for $PrO_{0.5}F_{0.5}BiS_2$. Lower inset shows the plot of $C_p$/T versus $T^2$ and its linear fitting. The plot for $C_p$ /T versus T is shown in the upper inset of the figure.

**Figure 1**

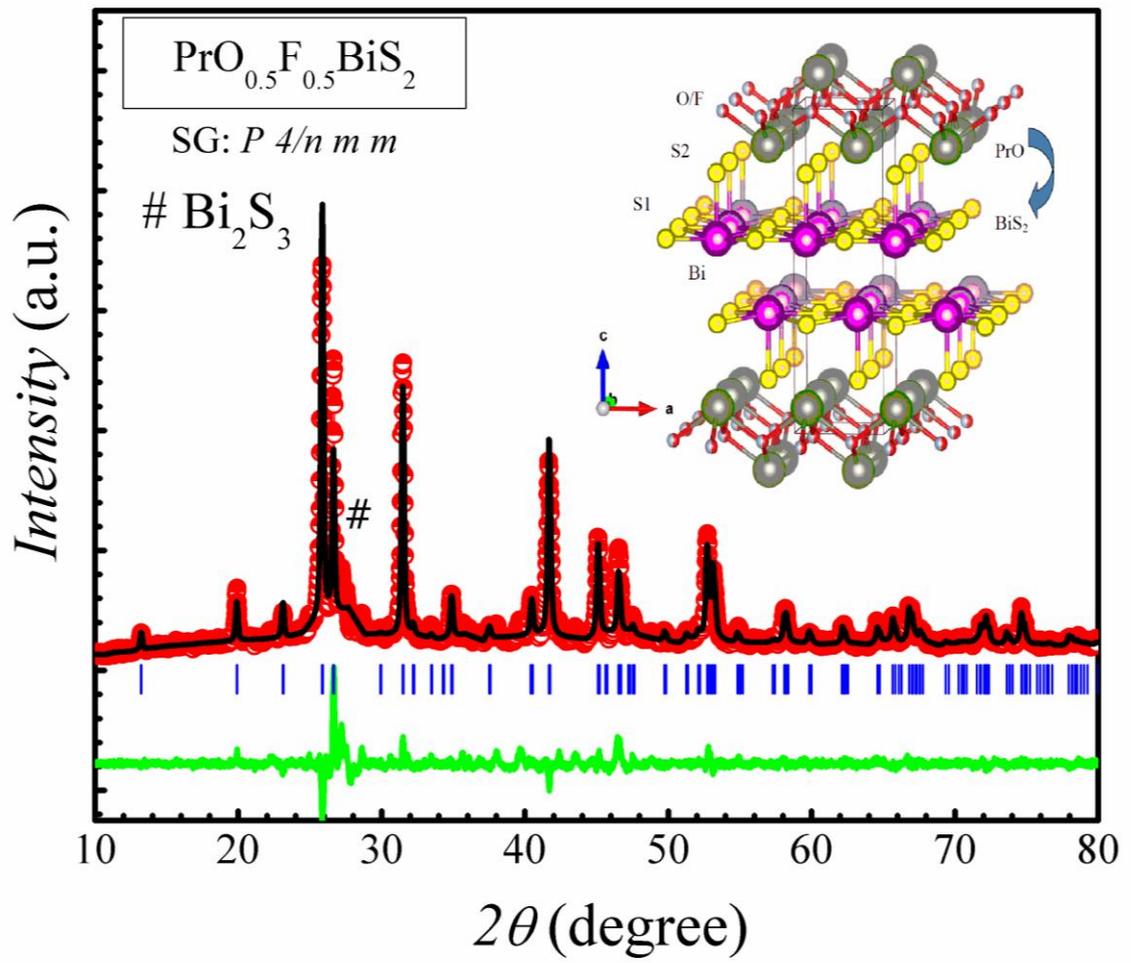

**Figure 2**

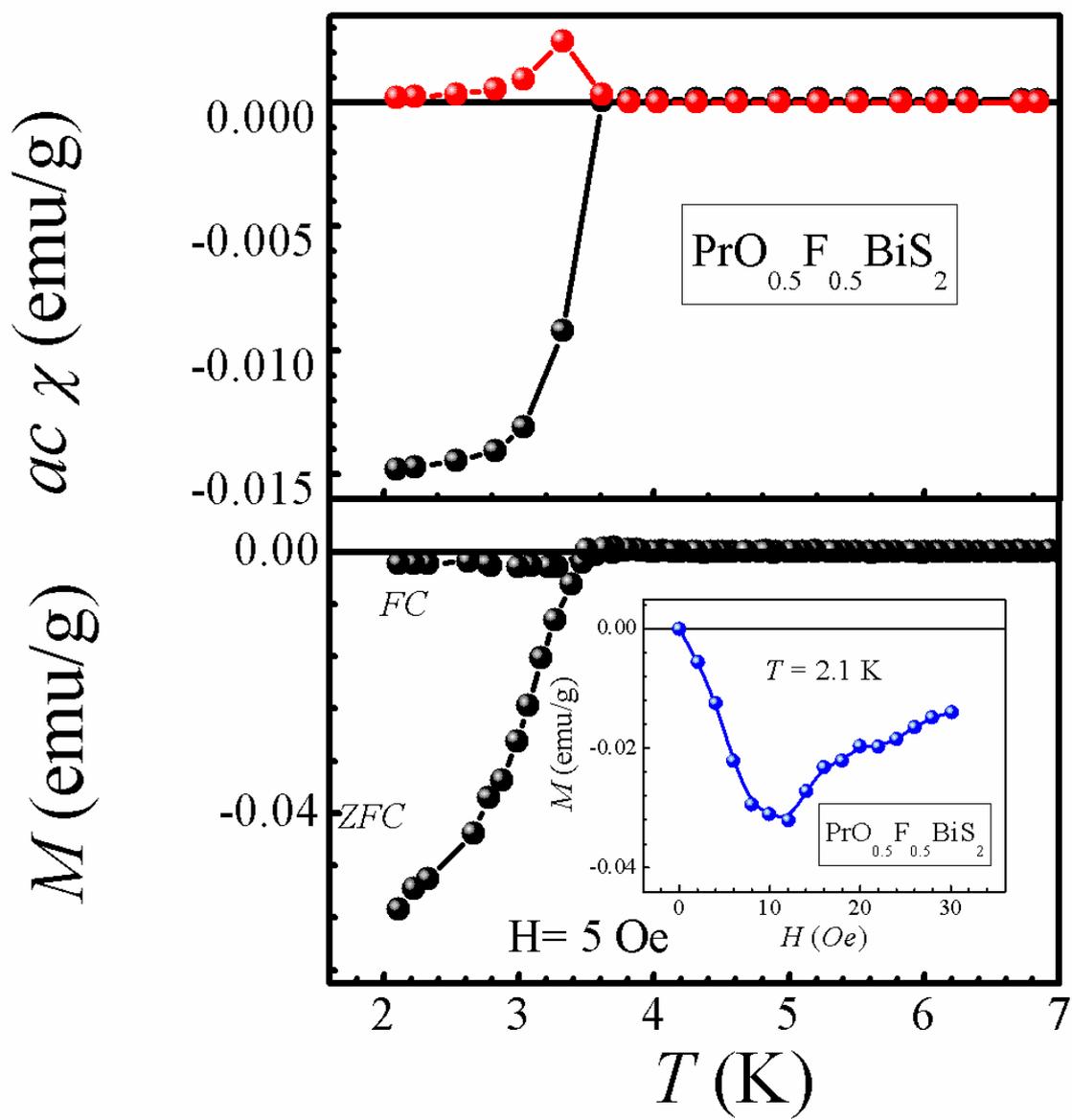

**Figure 3**

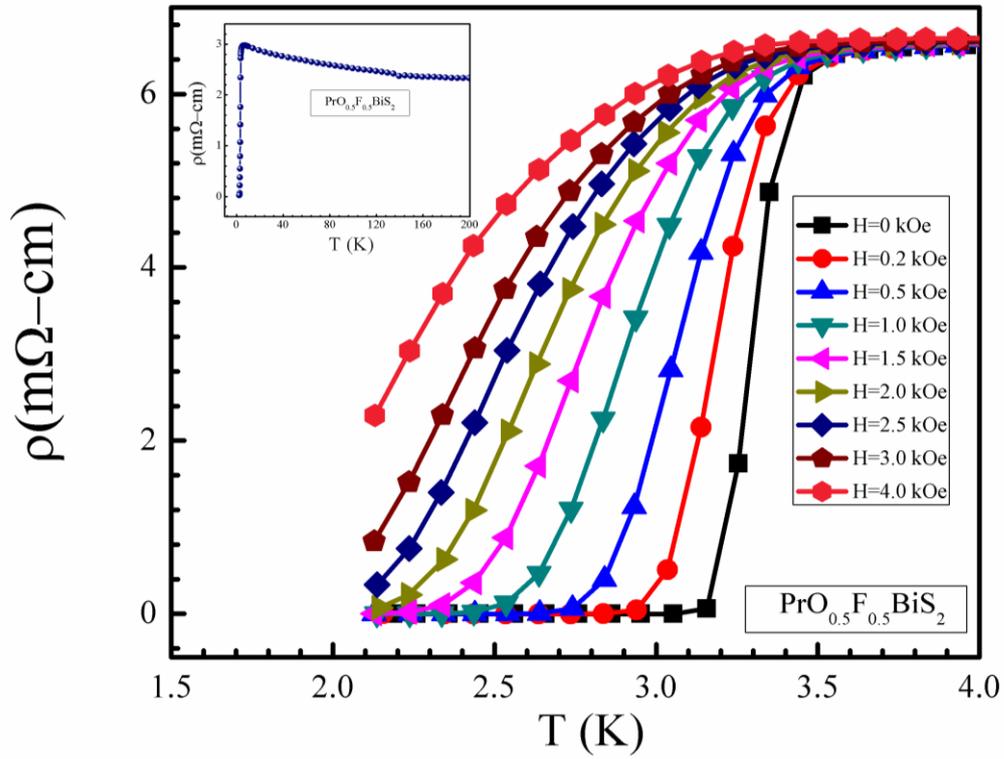

**Figure 4**

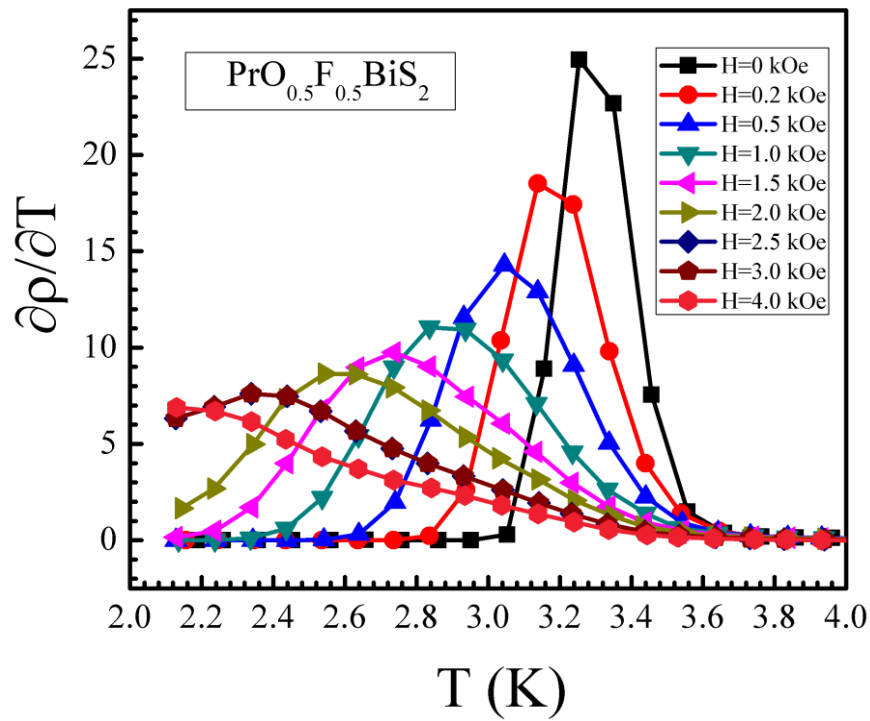

**Figure 5**

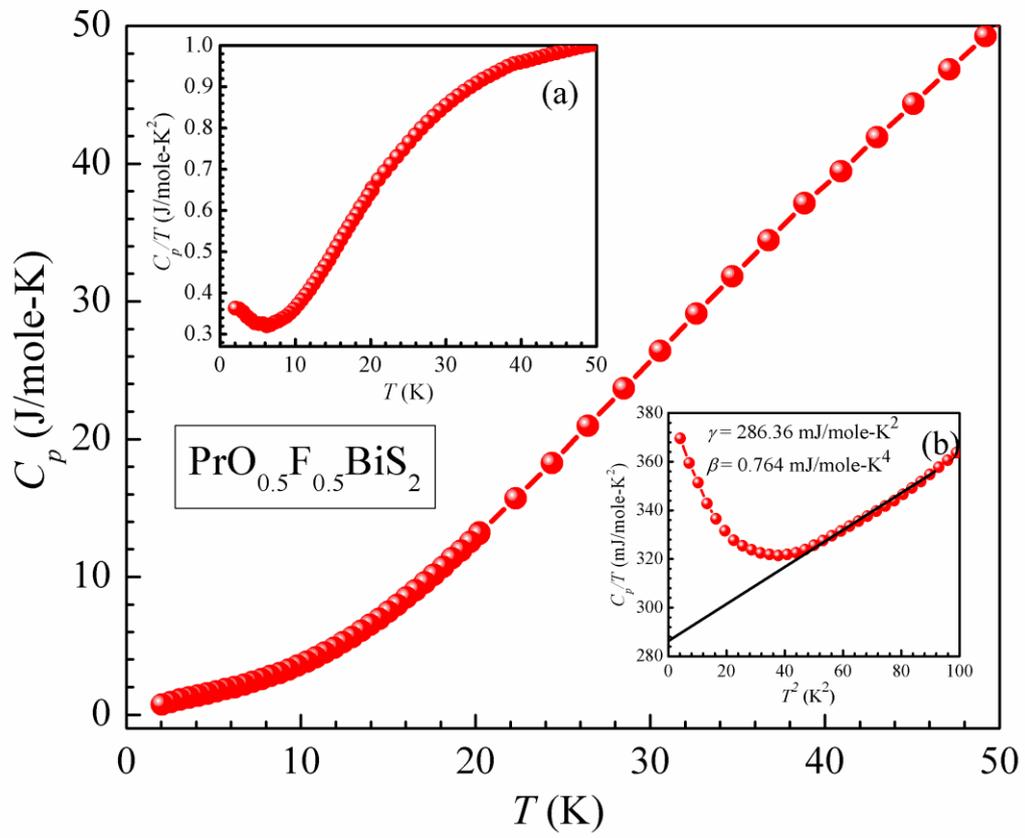